\documentclass[graybox]{svmult}

\usepackage{mathptmx}       
\usepackage{helvet}         
\usepackage{courier}        
\usepackage{makeidx}         
\usepackage{graphicx}        
\usepackage[caption=false,font=footnotesize]{subfig}
\usepackage{multicol}        
\usepackage[bottom]{footmisc}

\usepackage[cmex10]{amsmath}
\usepackage{amssymb}
\usepackage{stmaryrd}

\DeclareMathOperator*{\argmax}{arg\,max}
\newcommand{\nomalgo}{DynLOCNeSs}

\usepackage{algorithm}
\usepackage{algorithmic}

\makeindex             

\begin{document}

\title*{Vertex-centred Method to Detect Overlapping Communities in Evolving Networks}
\author{Ma\"{e}l Canu, Marie-Jeanne Lesot and Adrien Revault d'Allonnes}
\institute{Ma\"{e}l CANU \email{mael.canu@lip6.fr}
\and Marie-Jeanne Lesot \email{marie-jeanne.lesot@lip6.fr}
\at Sorbonne Universit\'es, UPMC Univ Paris 06, UMR 7606, LIP6, F-75005, Paris, France
\at CNRS, UMR 7606, LIP6, F-75005, Paris, France
\and Adrien Revault d'Allonnes \email{allonnes@ai.univ-paris8.fr}
\at Universit\'e Paris 8, EA 4383, LIASD, F-93526, Saint-Denis, France
}
%
%
\maketitle

\abstract{Finding communities in evolving networks is a difficult task and raises issues different from the classic static detection case. We introduce an approach based on the recent vertex-centred paradigm. The proposed algorithm, named \nomalgo{}, detects communities by scanning and evaluating each vertex neighbourhood by means of a preference measure, using these preferences to handle community changes. We also introduce a new vertex neighbourhood preference measure, CWCN, more efficient than current existing ones in the considered context. Experimental results show the relevance of this measure and the ability of the proposed approach to detect classical community evolution patterns such as grow-shrink and merge-split.
}

\section{Introduction}
%

A main task in computational network analysis is community detection, that consists in identifying denser subnetworks related to a specific role (eg. common interests in social networks, groups of interacting proteins in biological networks...) Though there is no universal definition for community, many have been proposed: intuitively, a community is a group of entities whose members have more relations between them than with the rest of the network. Many definitions and methods exist and keep being proposed~\cite{fortunato_community_2009,bedi_community_2016}.

Most community detection methods to date were designed to process static networks (see Section \ref{works}), however complex networks change over time and require methods able to take into account their dynamic (also referred to as temporal or evolutionary) dimension. It has been proved that straightforward use of static community detection algorithms at each time step (re-computation) is not relevant, in particular the output partition is not stable \cite{aynaud_static_2010}.

%
%

In this paper, we propose two contributions: first, an event-based detection algorithm relying on a vertex-centred process allowing a fast computation and a decentralised implementation, as well as a preference measure, \textbf{C}ommunity-based \textbf{W}eighted \textbf{C}ommon \textbf{N}eighbours (\textit{CWCN}) used in the vertex-centred process and more efficient than existing measures in the considered context.

The rest of this paper is organised as follows. Section \ref{works} presents related works about static and dynamic community detection methods. Section \ref{method} describes the principles of the proposed method \nomalgo{}, and introduces the vertex neighborhood measure CWCN. Experimental results to assess the ability of the method to capture simple network dynamics are provided in Section \ref{expes}. 

\section{Related Works}
\label{works}
We first present here static and dynamic community detection methods relevant to this paper. Other classic methods are reviewed in \cite{fortunato_community_2009,xie_overlapping_2013,bedi_community_2016}. Then, we review an approach more related to the proposed method: the vertex-centred paradigm.

\runinhead{Static Paradigms} Numerous static community detection approaches exist in the literature. They can be generic graph partitioning algorithms or take into account typical characteristics of the type of network they are designed for, such as power-law degree and small world effect in the case of social networks.

The main community detection method family is criterion optimisation. A global or local criterion measuring the quality of a graph partition into communities, such as the well-known modularity \cite{fortunato_community_2009}, is optimised through several iterations of an algorithm loop until convergence. Many existing criteria yield good quality partition (compared to a ground truth for example), but suffer from different drawbacks such as being subject to local extremum or resolution limit \cite{fortunato_resolution_2007}. This kind of method is also known to be time-consuming \cite{fortunato_community_2009}.

More recently, label propagation methods \cite{raghavan_near_2007,gregory_finding_2010,xie_overlapping_2013} offer a decentralised alternative. They rely on propagation of a node identifier (so-called ``label") from each vertex to every other in the network. However, despite being fast and suitable for detection in a decentralised environment, they have been found not to be stable as well \cite{leung_towards_2009,rezaei_near_2015}. Moreover, they make massive use of propagation and can overflood the network with unnecessary traffic, especially in a decentralised environment.

\runinhead{Dynamic Paradigms} The changeover from the static to the dynamic case is not easy. In particular, it depends on hypothesis about the graph evolution model. The most widespread considers a dynamic graph as a collection of static graphs, discretising the dynamic aspect with one graph instance per time step. Naive static detection on each time step, named static re-computation, has quickly been found to be unstable \cite{aynaud_static_2010}, especially when using optimisation methods, because the identified community structure varies too much, unrelatedly to the community evolution. For example, a good modularity value can be achieved on several very different community partitions of the same graph. To address this issue, concepts like \textit{temporal smoothness} introduced by Chakrabarti for evolutionary clustering were integrated~\cite{chakrabarti_evolutionary_2006}.

But even more than in the static case, taking into account the nature of the considered networks and the dynamics they are subject to is essential to design efficient methods \cite{lin_facetnet:_2008}. In this context, decentralised methods adaptation to process the dynamic case have been found to offer good performance, in terms of partition quality as well as computational efficiency, also offering the advantage to be easily implemented in parallel frameworks, as it is the case for label propagation \cite{leung_towards_2009,clementi_distributed_2014}. It is also very popular for applications in specific environments such as small decentralised mobile networks such as Pocket Switched Networks (PSN), for which community detection helps to improve network discovery and information routing \cite{hui_distributed_2007,orlinski_rise_2013}.
%
%


\runinhead{Vertex-centred Methods}

\label{vertex-centred}
Finally, vertex-centred approaches have gained popularity as a promising new community detection method family. They rely on the principle that some vertices in the network are ``leaders" or ``seeds" and the rest are followers \cite{riedy_detecting_2011}. Communities are formed by gathering followers around leaders, like in the \textit{Top-Leaders} approach \cite{rabbany_top_2010}. Although this method is more related to $k$-means clustering (re-allocating the leaders) than to a true leader-follower design, the introduced idea of expanding communities around leaders considering the potential \textit{preference} of a follower vertex (resp. a group of follower vertices) to join a leader vertex has been exploited by numerous algorithms. \textit{YASCA} \cite{kanawati_yasca:_2014} greedily expands communities around seeds and gather communities using ensemble clustering. \textit{LICOD} \cite{yakoubi_licod:_2014} starts with a careful selection of leaders before computing ranked community membership for each follower, then adjusting preferences and memberships using strategies borrowed from social choice theories until stabilisation. \textit{EMc} and \textit{PGDc} \cite{van_laarhoven_local_2016} locally expand around seed via EM or Projected Gradient Descent algorithm, using conductance to delimit communities. Canu et al. \cite{canu_fast_2015} consider each vertex as a potential leader and build preference dependencies allowing to form communities. True leaders are the core of the dependencies were the rest can be considered as followers.


Vertex-centred methods have also attracted attention to develop new dynamic community detection algorithms: for instance \textit{Evo-Leaders}, an adaptation of Top-Leaders \cite{gao_evolutionary_2016}, \textit{mux-LICOD}, an adaptation of LICOD for multiplex networks enabling use on evolving networks \cite{hmimida_community_2015}, OLEM/OLTM \cite{pan_online_2014} that locally optimises modularity and the original approach of \cite{zakrzewska_dynamic_2015} based on weighted-edge graphs, using weight update rules to cope with the dynamicity together with a fitness function to ensure partition quality.

We can also cite agent-based approaches like \textit{iLCD} consider each vertex as an agent and apply dynamic evolution rules to simulate the community formation, yielding a community structure \cite{cazabet_simulate_2011}.

The major drawback with these algorithms is that they loose one of the initial benefits of the leader-based approach, i.e. lightness and flexibility. Built on top of Top-Leaders, Evo-Leaders \cite{gao_evolutionary_2016} adds a costly split-merge of community at each time step. mux-LICOD \cite{hmimida_community_2015} uses degree centrality and shortest path calculation to compare leaders and followers. Shortest path computation can be costly if used for each vertex to each potential leader. It also relies on an aggregation phase repeated until stabilisation, though experiments do not reveal whether the stabilisation is fast or not. Finally, Zakrzewska et al.'s method \cite{zakrzewska_dynamic_2015} relies on a fitness function and a set of ad-hoc update rules and pruning over updates. It is hard to know however how efficient this policy is, as the experiments proposed by the authors are limited to a comparison with re-computation of the static counterpart. While faster than static re-computation (which is generally expected for specifically dynamic-addressed algorithms), the proposed $F$-score comparison with the set of static re-computed instances is not meaningful, as static re-computation has been proved to give unstable results \cite{aynaud_static_2010}.
%
%

%
%
%

\section{Proposed Approach}
\label{method}
This section describes the proposed approach, after defining the considered dynamicity model. We sketch the principles of the proposed method and describe in details the algorithm, which requires a vertex neighbourhood preference measure. We discuss such  preference measures and introduce a new one, \textit{CWCN}.

\subsection{Principles}
In the following, $G = (V, E)$ denotes an undirected graph, $\Gamma(v)$ for $v \in V$, the set of $v$'s neighbours and $d_v$ the degree of $v$. $C$ denotes the set of detected communities and $C(v)$ the community of $v$.
$S \subset V$ is the leader set, of all vertices being a leader for at least one other vertex. Each leader $s \in S$ has a set of followers $F(s) \subset V$. Alternatively, a follower $f$ has a set of preferred leaders, denoted $L(f) \subset V$. Preference measures between two vertices are denoted here using a preference function $\sigma: V \times V \rightarrow \mathbb{R}^+$.

\runinhead{Dynamicity}
We call \textit{time step} $t_i, i \in \mathbb{N}$ a date corresponding to a given state of the graph~$G$. The next time step $t_{i+1}$ occurs when at least an edge changes (appears or disappears).
The vertex event are treated as consequences of the edge moves: a vertex addition is captured as a new edge connecting a formerly isolated vertex. A vertex removal is captured in the same way, as the deletion of the last edge connecting this vertex to the rest of the graph. This constant vertex set model is widely used \cite{granell_benchmark_2015}.

We denote $G_i = (V, E_i)$ the state of $G$ 
and $C_i$ the state of communities at time $t_i$, eg. $G_0$ is the initial graph at $t_0$. Note that
the time interval $|t_i - t_{i-1}|$ is not necessarily constant.

\subsection{Proposed Algorithm: \nomalgo{}}
We propose \nomalgo{} (\textbf{Dyn}amic \textbf{L}ocating of \textbf{O}verlapping \textbf{C}ommunities in \textbf{Ne}work \textbf{S}tructure\textbf{s}), a vertex-centred approach to detect communities in dynamic graphs, more precisely a leader-based approach using a vertex neighbourhood preference measure. The idea is to change from a batch to an event-based detection and modification process, and to perform the detection with as little as possible re-computation. Each vertex must determine whether it should change its leader. If so, it may also change community.

The proposed method takes as input an initial graph, $G_0$, along with initial community structure $C_0$ and leader set $S_0$, and only deals with the detection over time. These initial states can be computed using any leader-based method (see Section \ref{vertex-centred}). The implementation presented here uses an approach in which each vertex $v \in V$ is considered as a potential leader and evaluates its neighbourhood, like \textit{iLCD} \cite{cazabet_simulate_2011} or Canu et al. \cite{canu_fast_2015}. It has the advandage of not pre-selecting a set of leaders, thus not suffering from the bad seed selection issue.

The main part of the algorithm is the \textit{vertex update procedure} described in Algorithm \ref{algo:eval}). It is run when an edge (dis)appears, which is the only event considered here. The algorithm also relies on a times-step related vertex marking, which is used to identify whether the leaders or community must be re-computed. The marking is explained first, and then the vertex update procedure.

\runinhead{Marking}
A vertex is marked to signify it has changed community, and is meant to be seen only by the vertex neighbours. The marks made at $t_i$ are visible at time $t_{i+1}$. Vertices having a marked vertex in their leader set will reconsider their community membership. This marking is the way to accelerate changes propagation through the graph, because a community change for a vertex increases the probability of one of its neighbours to change community too.

\runinhead{Vertex Update Procedure}
This procedure is run for a vertex $v$ only if a change occured in its neighbourhood, the only possibility that may lead to a community change for $v$.
In this case, at time $t_i$, each vertex $v$ locally computes all the preferences between itself and its neighbours, ie. all the $\sigma(v, v')$ for all $v' \in \Gamma(v)$. Because of the neighbourhood change, a leader could have disappeared or a new one appeared. If the new preferences values imply an actual change in $L(v)$, the community of $v$ is also re-evaluated. If that results in $C(v)$ changing, then $v$ \textit{marks} itself as previously stated.

\runinhead{Flexibility and Local Computation.} The proposed algorithm only uses local computations from each vertex, thus keeping the vertex-centred methods flexibility advantage. This allows an easy decentralised implementation in Pregel-like frameworks (see \cite{mccune_thinking_2015}): the vertex program is simple to write and few informations are susceptible to be shared between parallel processes.
	


%

\begin{algorithm}[t]
\caption{Vertex Update Procedure for time step $t_i$}
\label{algo:eval}
\begin{algorithmic}[1]
	\REQUIRE ~\\
	$v \in V$, a vertex\\
	$\Gamma_i(v)$, its neighbours at time $t_i$
	\ENSURE ~\\$C_i(v)$, updated community for $v$
	
	\IF{$\Gamma_i(v) \neq \Gamma_{i-1}(v)$}
		\STATE recompute $v$'s preferred leaders : $L(v) \leftarrow \argmax_{u \in \Gamma_i(v)} \sigma(v, u)$
		\IF{$L(v)$ changes \textbf{or} any $u \in L(v)$ is marked}
			\STATE $C_i(v) \leftarrow $ most frequent community among $L(v)$
			\IF{$C_i(v) \neq C_{i-1}(v)$}
				\STATE mark each $v$ for time $t_i$
			\ENDIF
		\ENDIF
	\ENDIF
\end{algorithmic}
\end{algorithm}

\subsection{Preference Measures}
\label{preference}
The proposed method relies on a vertex neighbourhood preference measure $\sigma: V \times V \rightarrow \mathbb{R}^+$, evaluating at which point a vertex $v \in V$ is close to a given neighbour $u \in \Gamma(v)$. It must reflect a closeness or attraction dynamics at work in the graph. For example, in a social network, $\sigma(v, u)$ must account for the friendship level of $v$ towards $u$. Such closeness often relies on the quantity of common neighbours between $u$ and $v$. The measures presented below make use of these quantities.

We review here three measures as presented in \cite{cohen_survey_2012} (Section 2.2), and propose a new proposed measure \textbf{C}ommunity-based \textbf{W}eighted \textbf{C}ommon \textbf{N}eighbours (CWCN), taking into account known information community. Section \ref{expes} presents results of the algorithm implementing each of these measures. The mathematical expression is given for each measure for any $u, v \in V$.

\textit{Jaccard} coefficient of neighbours is an adaptation of the well-known \textit{Jaccard Index} for neighbour vertices in a graph, and compares the number of common neighbours to the total number of neighbours of both $u$ and $v$. It is defined as follows:
\begin{equation}
	\sigma_{Jac}(u, v) = \frac{|\Gamma(u) \cap \Gamma(v)|}{|\Gamma(u) \cup \Gamma(v)|}
\end{equation}

\textit{Adamic-Adar} is an adaptation of the eponymous measure used for web search and link prediction. It sums the number of common neighbours between $u$ and $v$, using a logarithmic function that gives more importance to ``rarer" features, here to less connected neighbours. It is defined as follows: 
\begin{equation}
	\sigma_{AA}(u, v) = \sum_{w \in \Gamma(u) \cap \Gamma(v)} \frac{1}{\log(|\Gamma(w)|)}
\end{equation}

The \textit{Preferential Attachment}  measure is based on the eponymous concept popularised by Barab\'{a}si and Albert \cite{barabasi_emergence_1999}: the tendency of entities having many connections to attract more new connections than weakly connected ones. It multiplies the neighbourhood sizes of $u$ and $v$, meaning that preference hugely depends on vertex degree. Using this measure results in large agglomerations of vertices around hubs. It is defined as follows:

\begin{equation}
	\sigma_{PA}(u, v) = |\Gamma(u)| \times |\Gamma(v)|
\end{equation}
\par

\runinhead{The proposed measure} \textit{Community-based Weighted Common Neighbours}, is a common neighbour measure weighted by the degree of the vertex being compared. While similar to the common neighbours $|\Gamma(u) \cap \Gamma(v)|$, the degree weighting scheme ``attracts'' a vertex much more toward high degree leaders and thus higher density areas in the graph, related to communities. This follows Barab\'{a}si \& Albert's preferential attachment principle \cite{barabasi_emergence_1999} but is less strong that the preferential attachment measure described above. It is given by:

\begin{equation}
	\sigma_{CWCN}(u, v) = |\Gamma(u) \cap \Gamma(v)| \times d_v
\end{equation}
\par





\section{Experiments}
\label{expes}

This section presents several experiments supporting the validity of the proposed method. It compares the effectiveness of various preference measures presented Section \ref{preference}. The goal of these experiments is to prove the ability of \nomalgo{} (together with an appropriate preference measure) to capture the dynamics of evolution of the network, and as such is done on small interpretable graphes, with experiments similar to \cite{granell_benchmark_2015}. The experiments on big graphs (data mining) are left to future works.


\subsection{Protocol}
\runinhead{Datasets.} We use artificial benchmark graphs to assess the properties and validity of the proposed algorithm. They are obtained using the generator proposed by Granell et al. \cite{granell_benchmark_2015}. It keeps the vertex set constant and uses two community evolution patterns: grow-shrink where some communities grow (gain vertices) while others shrink (lose vertices), and merge/split where merge and splits occur between communities. It can generate an evolving graph of controlled size and density after one or both patterns, together with the ground truth community structure. We specify for each experiment the benchmark parameters used to generate the graphs.
%
%
%

\runinhead{Evaluation Criteria.} We use the same criteria for partition comparison as in \cite{granell_benchmark_2015}: the classical information entropy-based measures \textit{Normalised Variation of Information} (NVI) and \textit{Normalised Mutual Information} (NMI), both bounded between $[0, 1]$. However, opposite to the NVI, a NMI value of 1 indicates that the two partitions contain the same information (identical) whereas 0 indicates that the partitions are totally dissimilar. A good community structure partition thus minimises the NVI and maximises the NMI. See the mathematical expressions in \cite{granell_benchmark_2015}.
%
%

We choose not to use the proposed windowed variant \cite{granell_benchmark_2015} as it does not bring significant benefit and it is difficult to interpret. As a matter of fact it requires to carefully select the time window value, leading to significantly impact the relative performance of two methods if improperly done.

\subsection{Preference Measure Comparison}
The first experiment is performed in order to compare the effect of the different preference measures exposed in Section \ref{preference}. We use here the classic planted bissection model \cite{clementi_distributed_2014,granell_benchmark_2015}. In this model, the graph is divided into two communities and the algorithm has to correctly classify each vertex as belonging to one or the other.

\begin{figure*}[t]
       \subfloat[Ground truth]{
       		\includegraphics[width=0.31\textwidth]{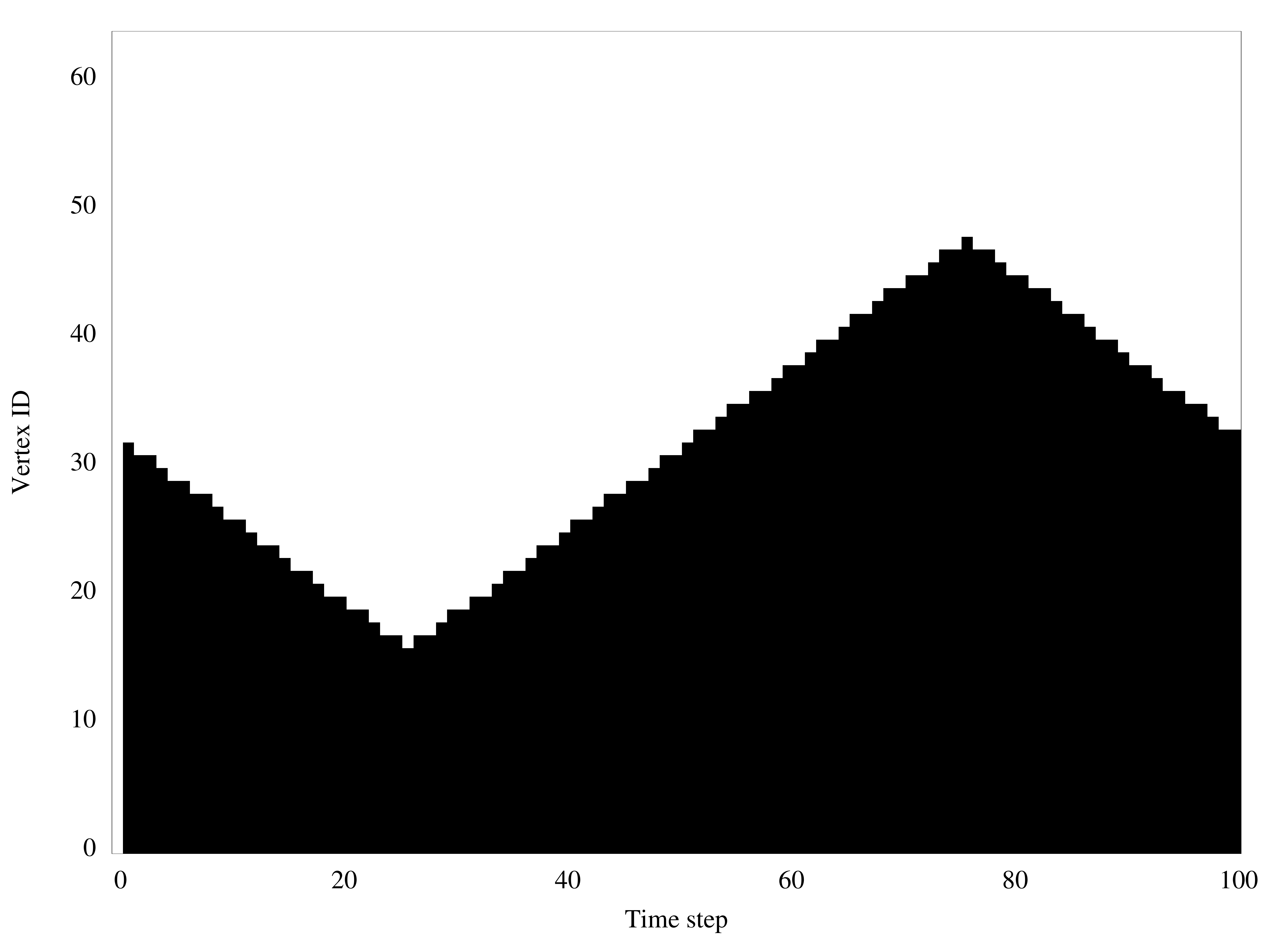}
       		\label{fig:grow_gt}
		}
		\vspace{0.15in}
       \subfloat[With $\sigma_{CWCN}$]{
       		\includegraphics[width=0.31\textwidth]{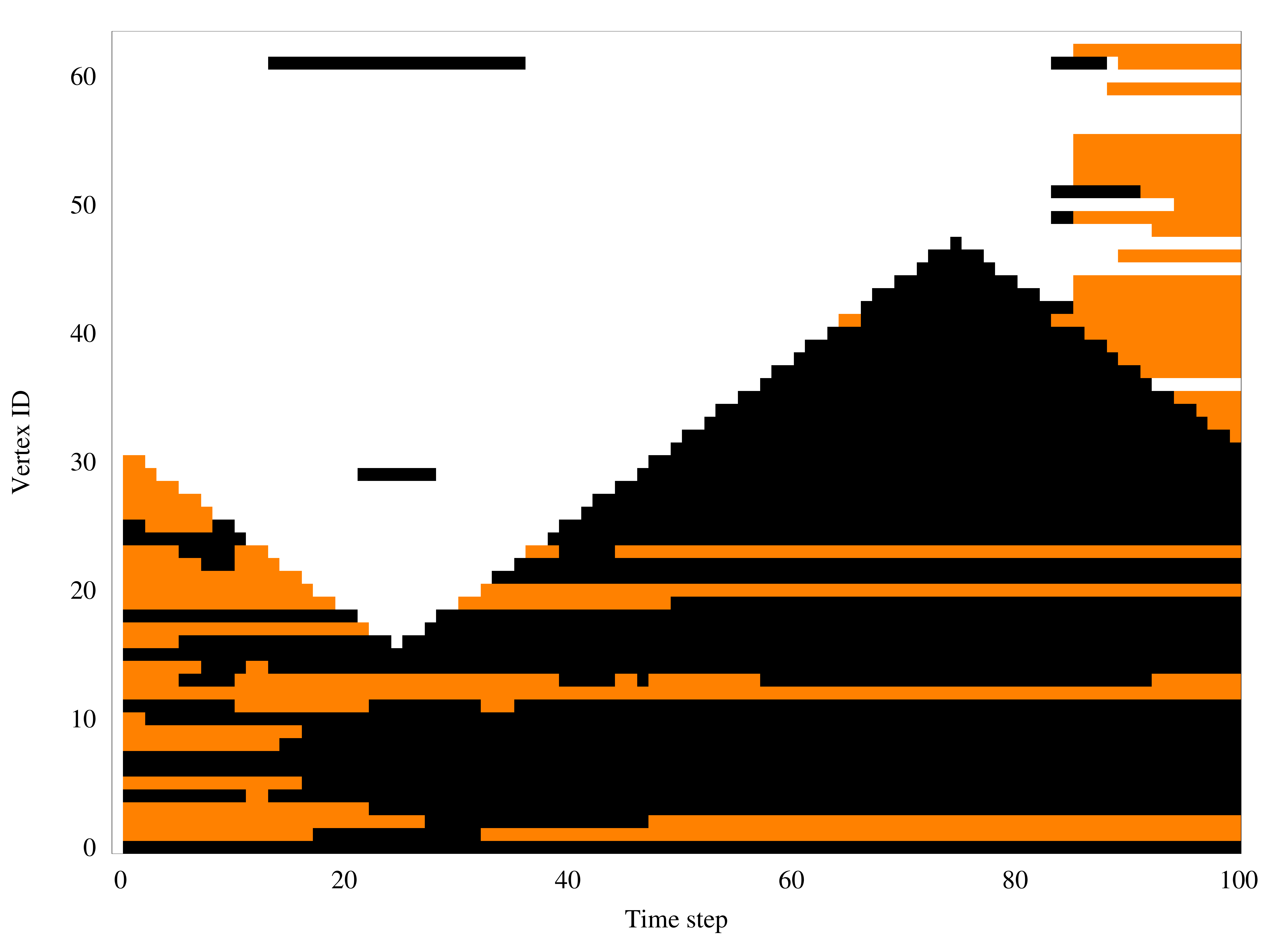}
		}
		\vspace{0.15in}
		\subfloat[With $\sigma_{Jac}$]{
       		\includegraphics[width=0.31\textwidth]{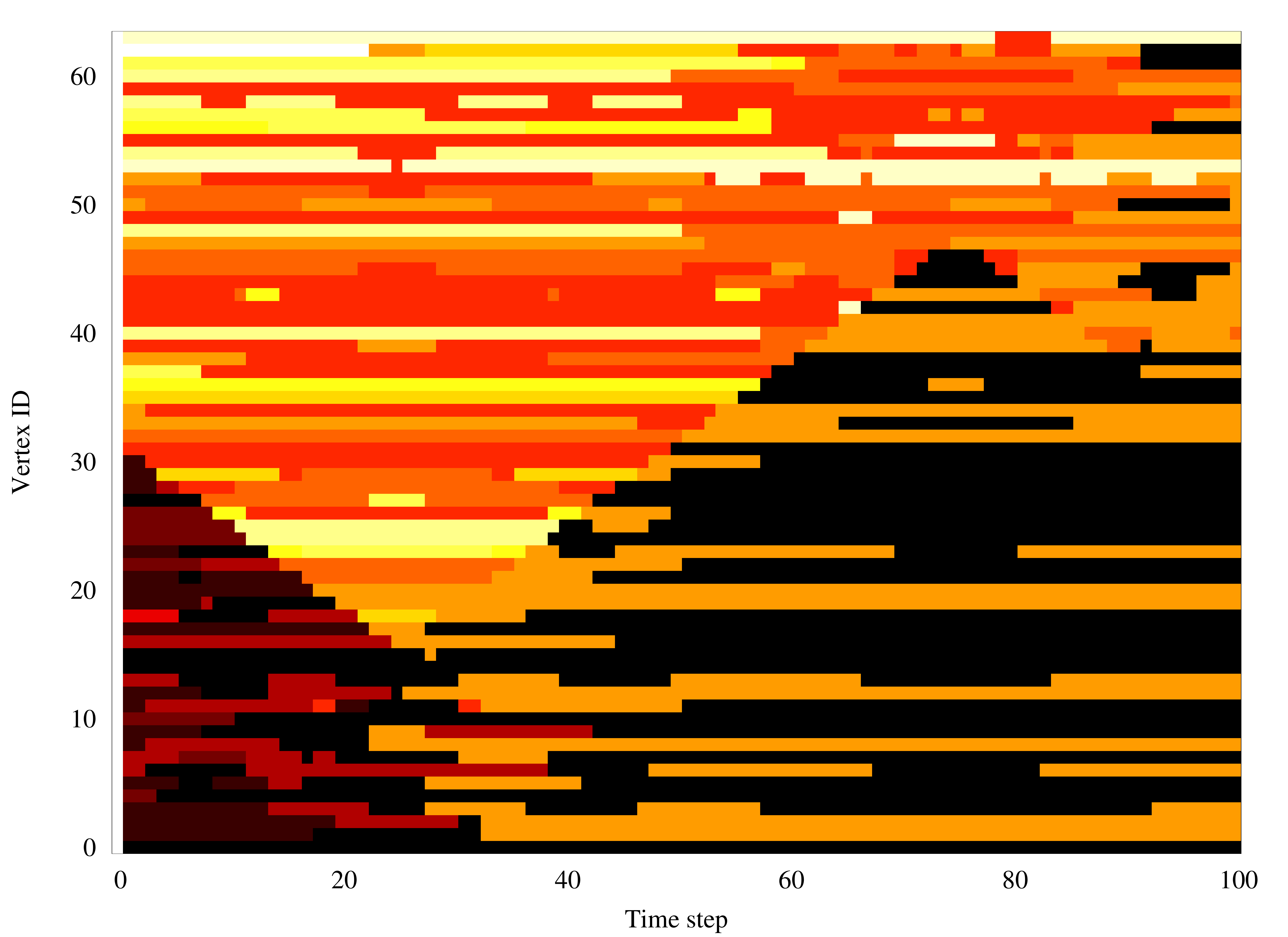}
		}
		\caption{(Colors online) Comparative visualisation of the community repartition between $\sigma_{Jac}$ and $\sigma{CWCN}$ evaluated on the grow-shrink process.} 
		\label{fig:colormaps_grow}
\end{figure*}

\begin{figure*}[t]
       \subfloat[Ground truth]{
       		\includegraphics[width=0.31\textwidth]{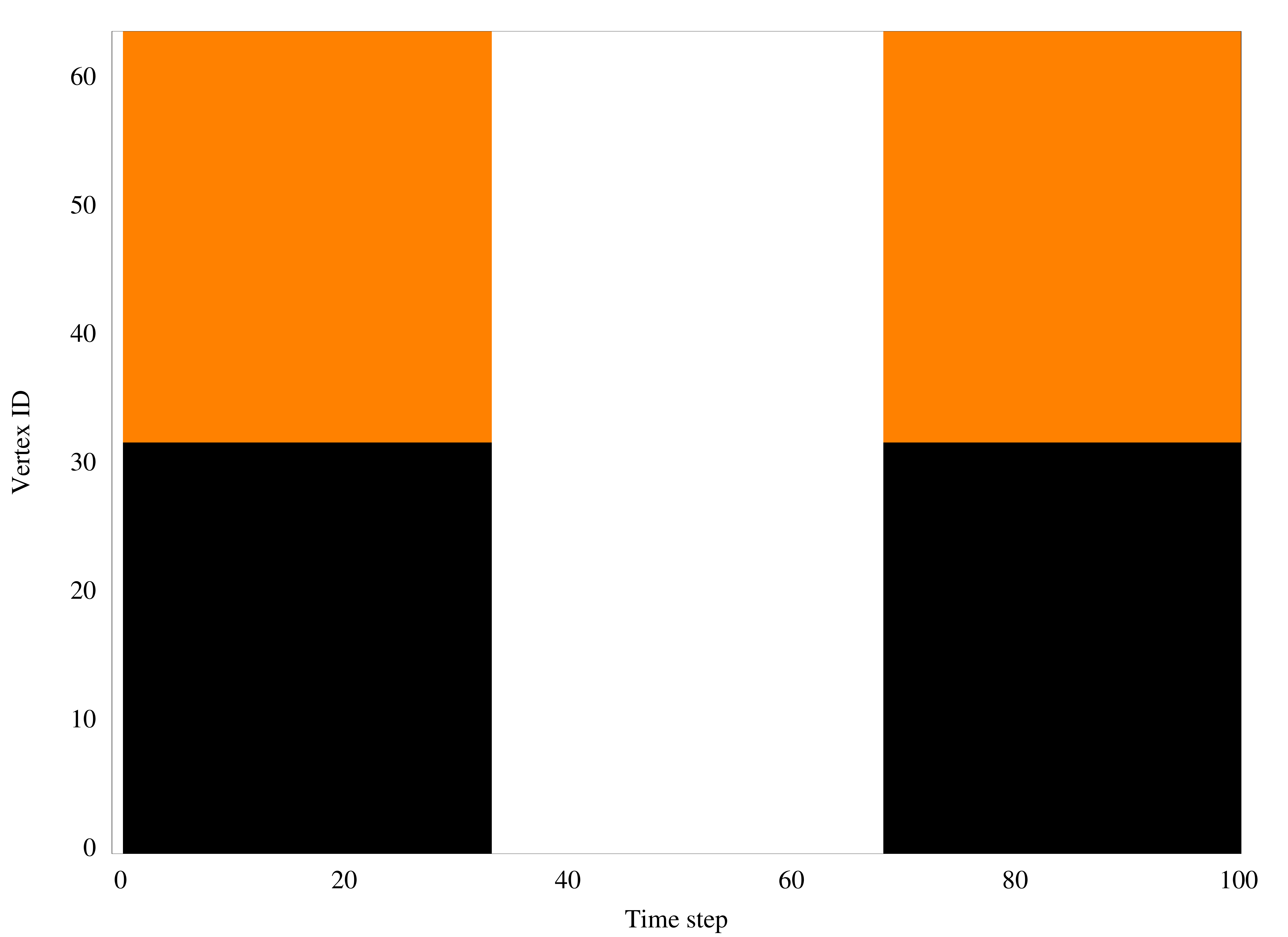}
       		\label{fig:merge_gt}
		}
		\vspace{0.15in}
       \subfloat[With $\sigma_{CWCN}$]{
       		\includegraphics[width=0.31\textwidth]{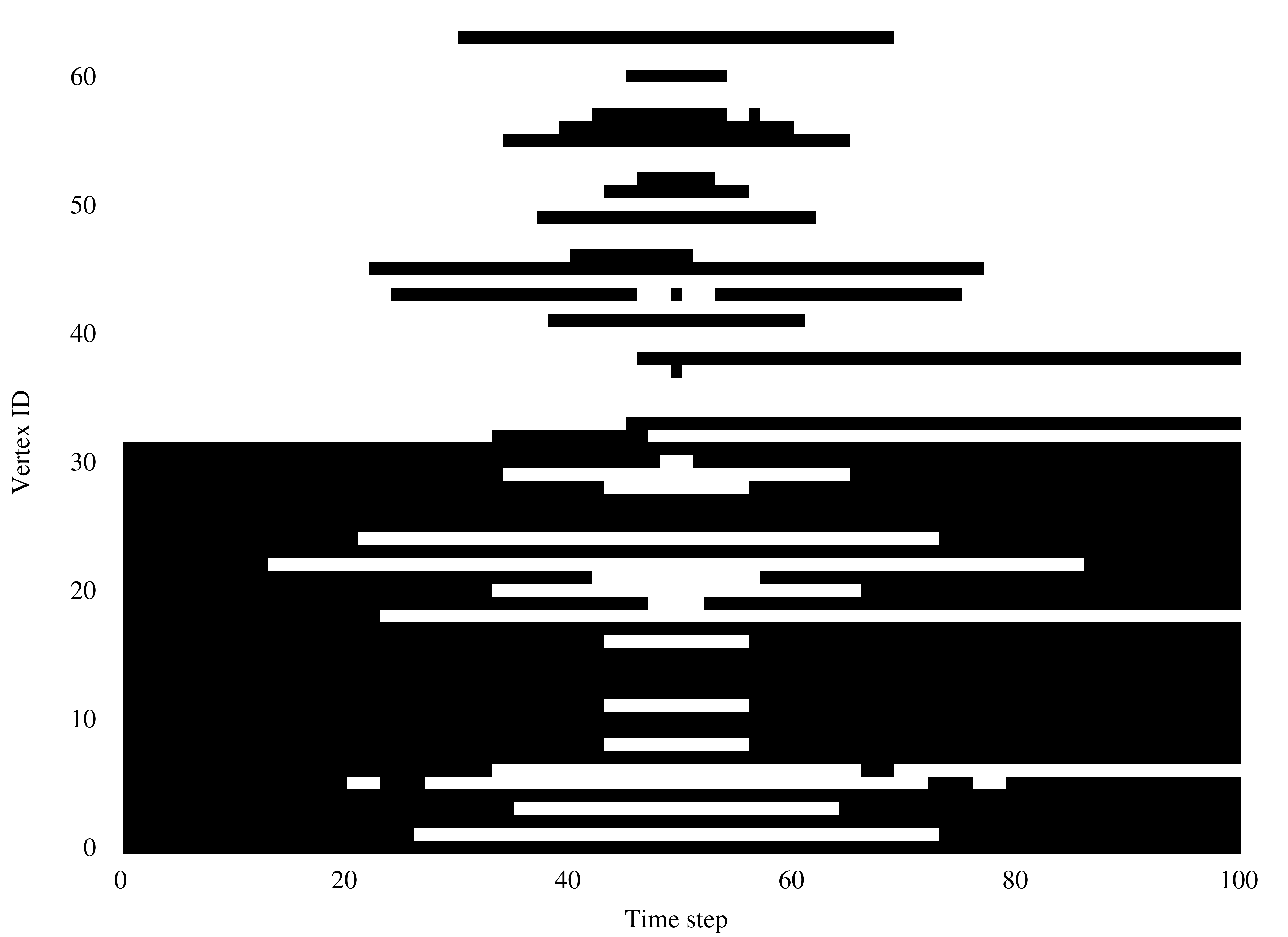}
		}
		\vspace{0.15in}
		\subfloat[With $\sigma_{Jac}$]{
       		\includegraphics[width=0.31\textwidth]{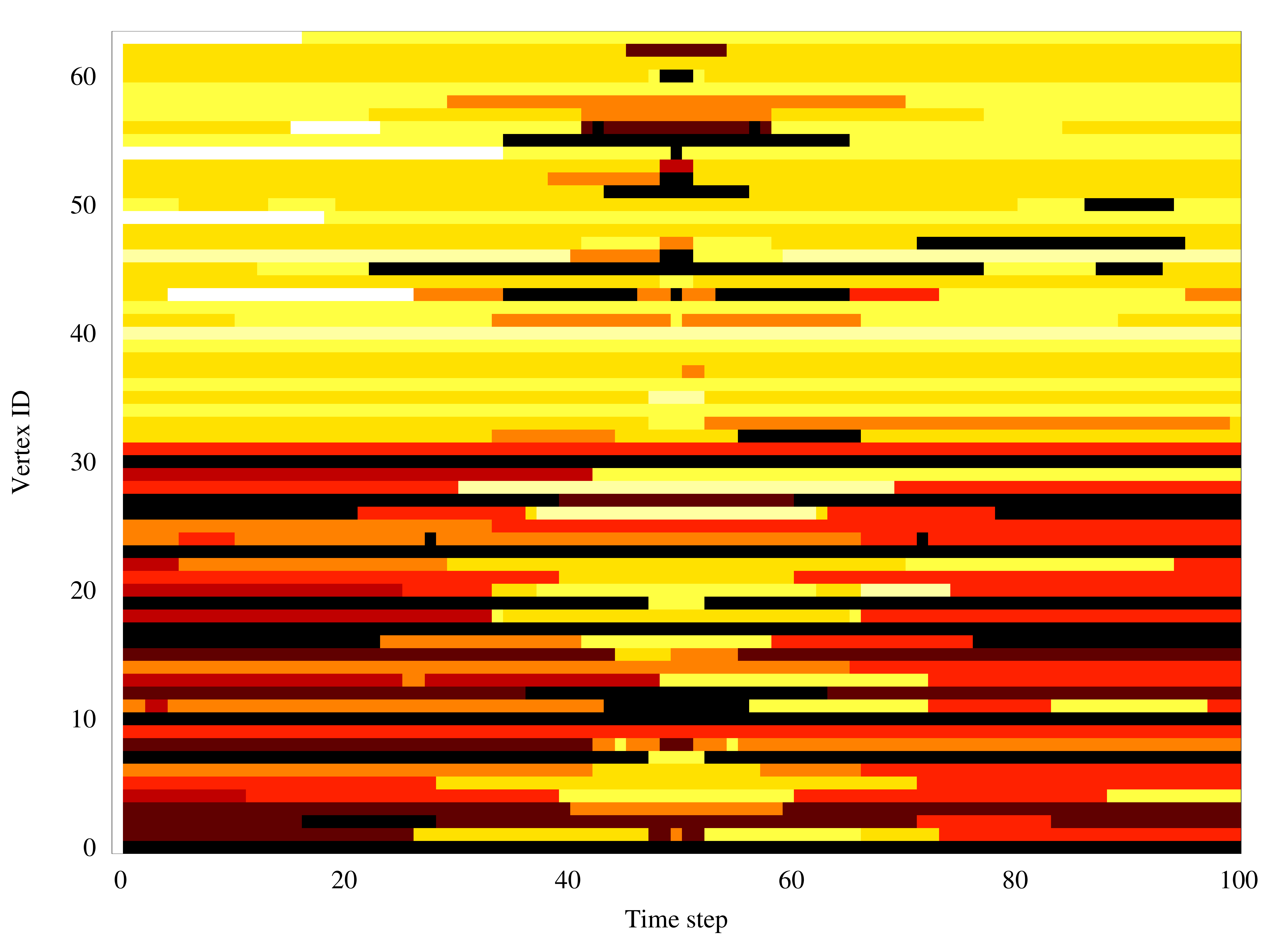}
		}
		\caption{(Colors online) Comparative visualisation of the community repartition between $\sigma_{Jac}$ and $\sigma_{CWCN}$ evaluated on the merge-split process.} 
		\label{fig:colormaps_merge}
\end{figure*}

The proposed algorithm is tested for each preference measure on two evolution patterns : grow-shrink and merge-split. For each pattern, 10 instances of a graph of $64$ vertices are generated, with intra-community density of 0.5 and inter-community density of 0.05, for 100 time steps. These parameters are those used in~\cite{granell_benchmark_2015}. The ground truth, shown on Fig. \ref{fig:grow_gt} and \ref{fig:merge_gt}, is thus made of 2 communities of 32 vertices each at $t_0$.


%
\begin{figure*}[t]
       \subfloat[NVI (to be minimised)]{
       		\includegraphics[width=0.48\textwidth]{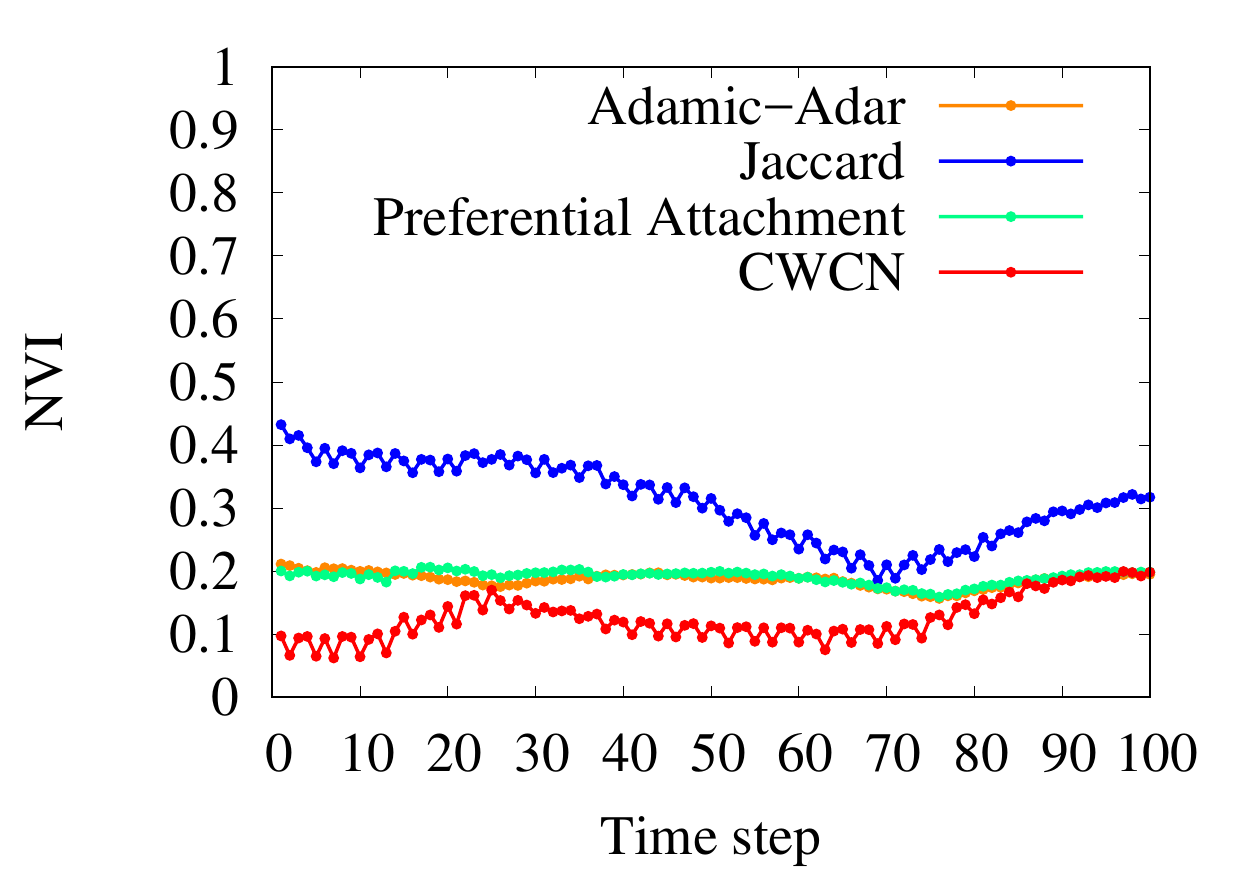}
		}
		\vspace{0.15in}
		\subfloat[NMI (to be maximised)]{
       		\includegraphics[width=0.48\textwidth]{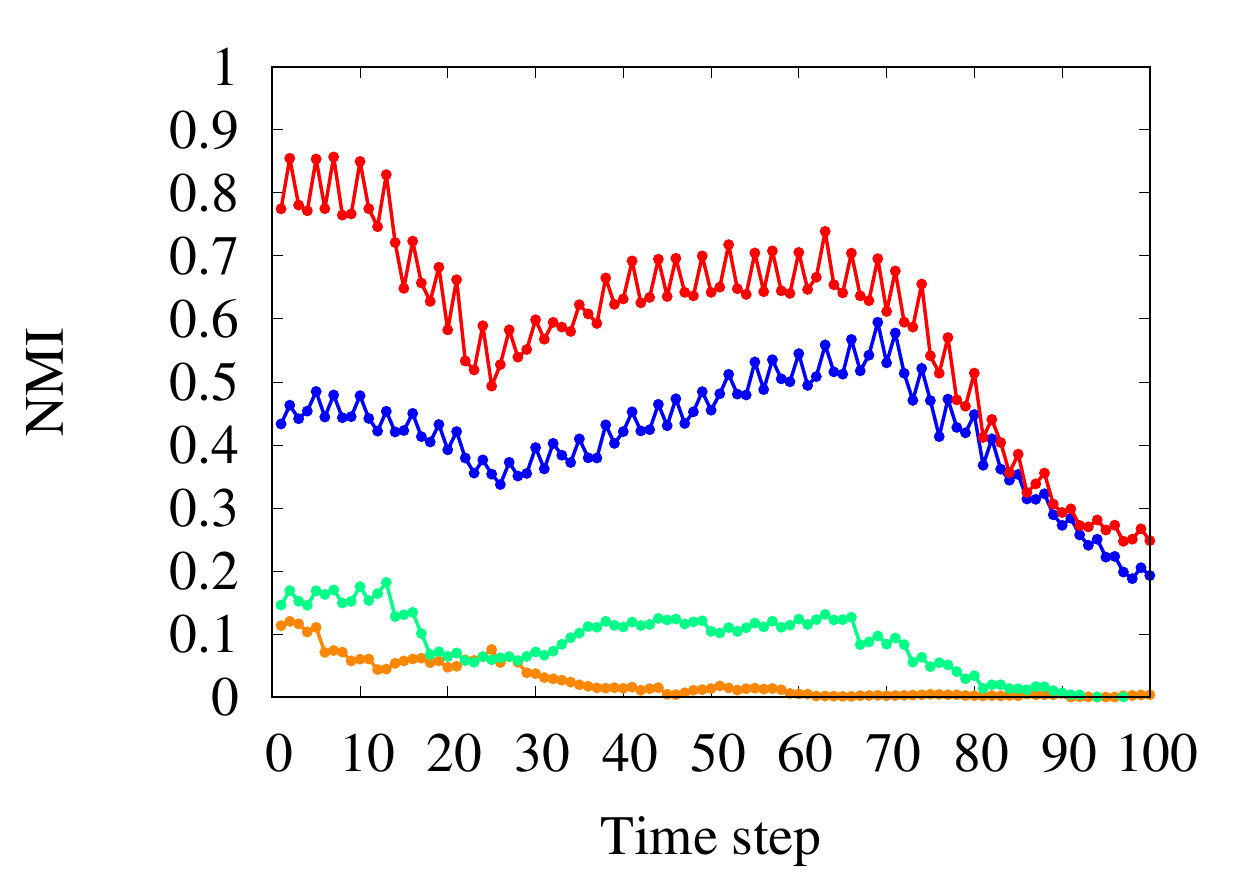}
		}
		\caption{Comparison for the grow-shrink pattern on 100 time steps}
		\label{fig:curv_grow}
\end{figure*}

Results for the Grow-Shrink pattern are presented
on Figures \ref{fig:curv_grow} and \ref{fig:colormaps_grow} (the measures not shown on Figure \ref{fig:colormaps_merge} produce only one community at each time step, therefore the colormap is all black) are the mean of NVI and NMI runs over the 10 graphs, and a colormap visualisation where each pixel color represents the community assignment of a vertex (id on the $y$ axis) at a given time step (on the $x$ axis). We can see that \nomalgo{} with $\sigma_{CWCN}$ globally detects the grow-shrink bissection pattern, except that a third community (orange) is identified. This community in fact replaces the black one at the beginning and the white one at the end: the method takes the grow-shrink evolution as a transfer between two communities via a third one, impacting NVI and NMI values. However, the clearly visible grow-shrink triangle shapes indicate that the evolution pattern has correctly been identified.

This is less obvious for the method with $\sigma_{Jac}$. It detects 14 communities and even if the triangle shape can be guessed there is a lot of noise and community misassignment.

The other two cases, $\sigma_{AA}$ and $\sigma_{PA}$, are not pictured because they assign every vertex to a single community, resulting in an entire black colormap.

\begin{figure*}[t]
       \subfloat[NVI (to be minimised)]{
       		\includegraphics[width=0.48\textwidth]{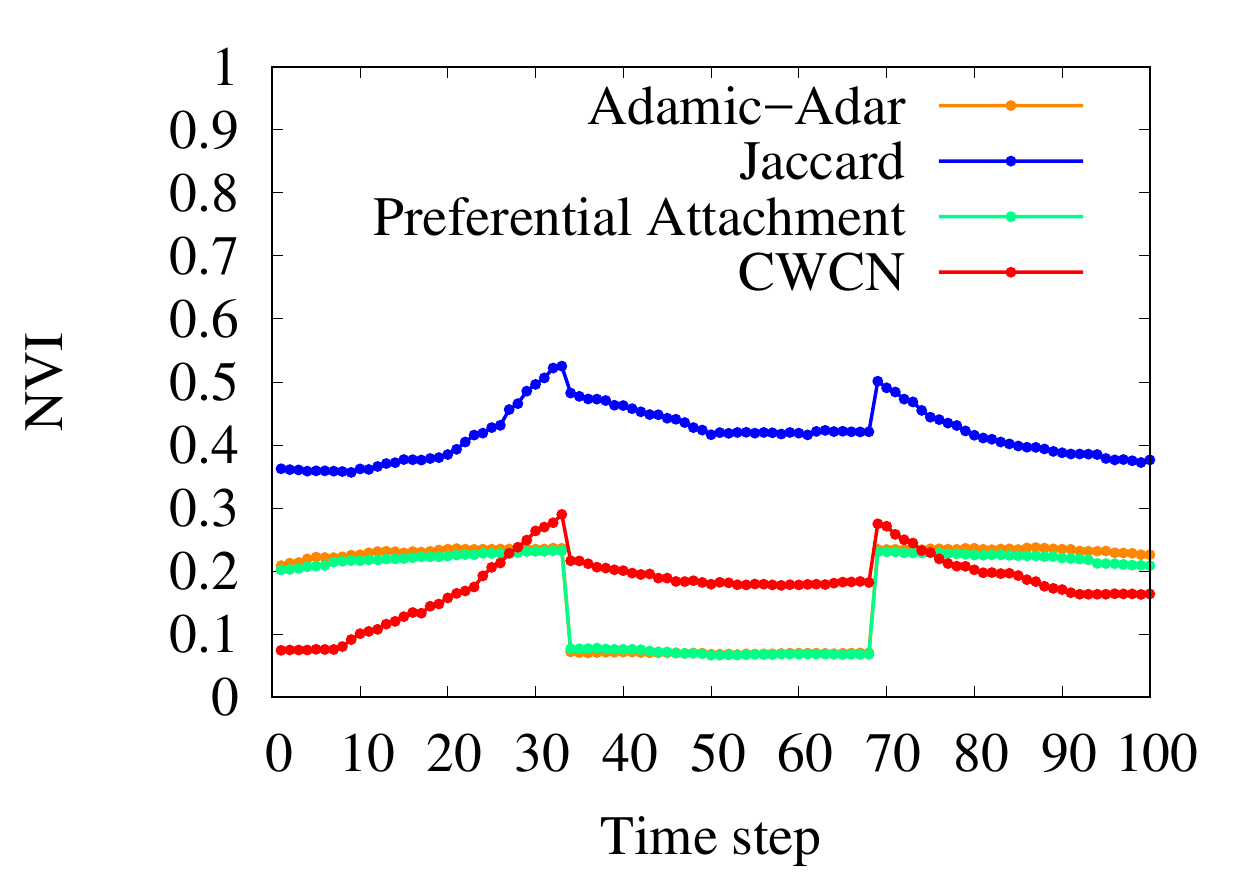}
		}
		\vspace{0.15in}
		\subfloat[NMI (to be maximised)]{
       		\includegraphics[width=0.48\textwidth]{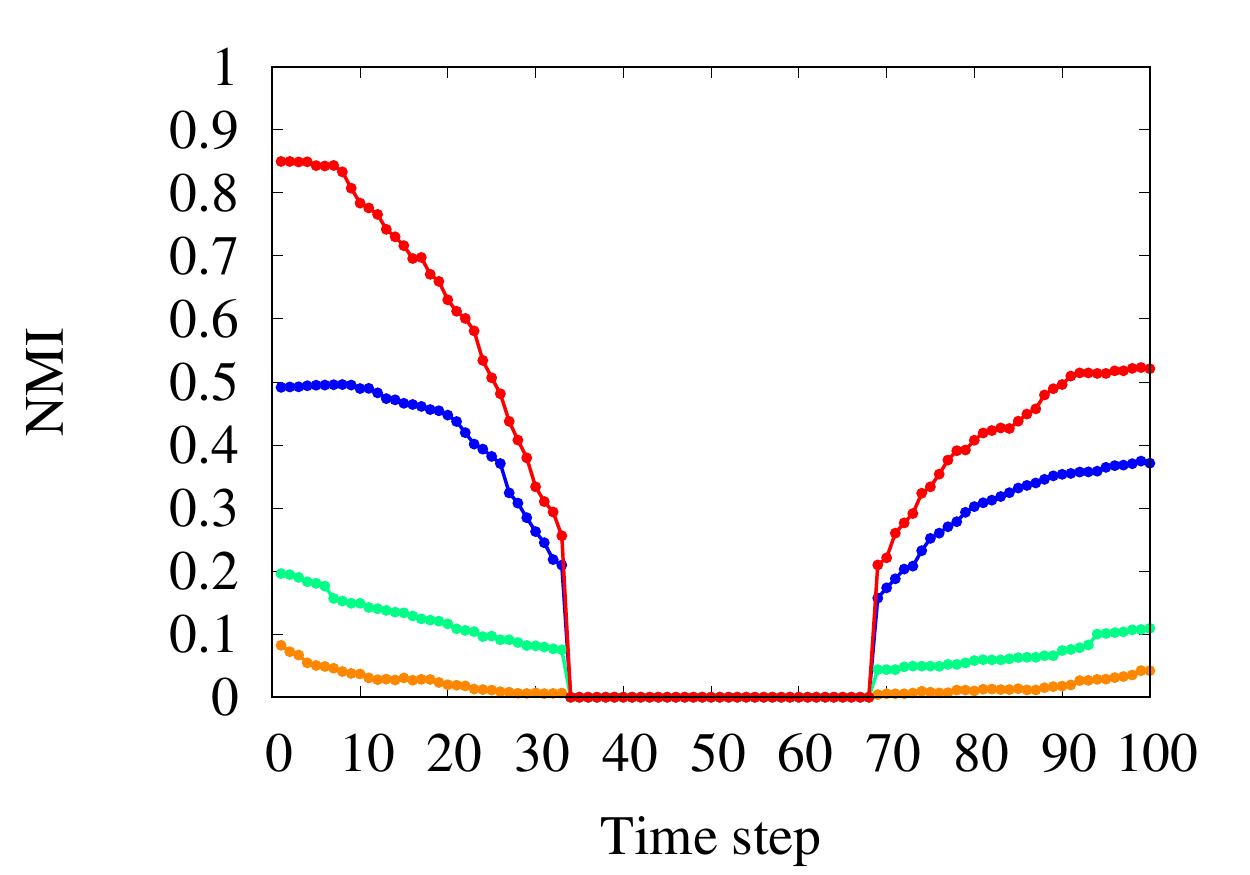}
		}
		\caption{Comparison for the merge-split pattern on 100 time steps}
		\label{fig:curv_merge}
\end{figure*}

The merge-split process is presented in Figures \ref{fig:curv_merge} (criteria) and \ref{fig:colormaps_merge} (visualisation). Again, the measures not shown on Figure \ref{fig:colormaps_merge} produce only one community at each time step, therefore the colormap is all black.

Merge-Split is less successfully detected. We notice that $\sigma_{CWCN}$ finds two communities where $\sigma_{Jac}$ finds ten, but the abrupt merge is not correctly identified, although the CWCN variant yields less noise than Jaccard one.

Let aside the merge, the CWCN variant nonetheless achieves better NMI and NVI than the other methods. The perfect NVI for $\sigma_{AA}$ and $\sigma_{PA}$ during the merge can be explained by the fact that both methods only detect one community at any time step. This is prejudiced when two communities exist, but it is correct during the merge. It is a side effect related to the chosen planted bissection, but inherently denotes a poor quality of detection for these two criteria.

\runinhead{Execution Time}
\begin{figure*}[t]
		\centering
		\includegraphics[width=0.70\textwidth]{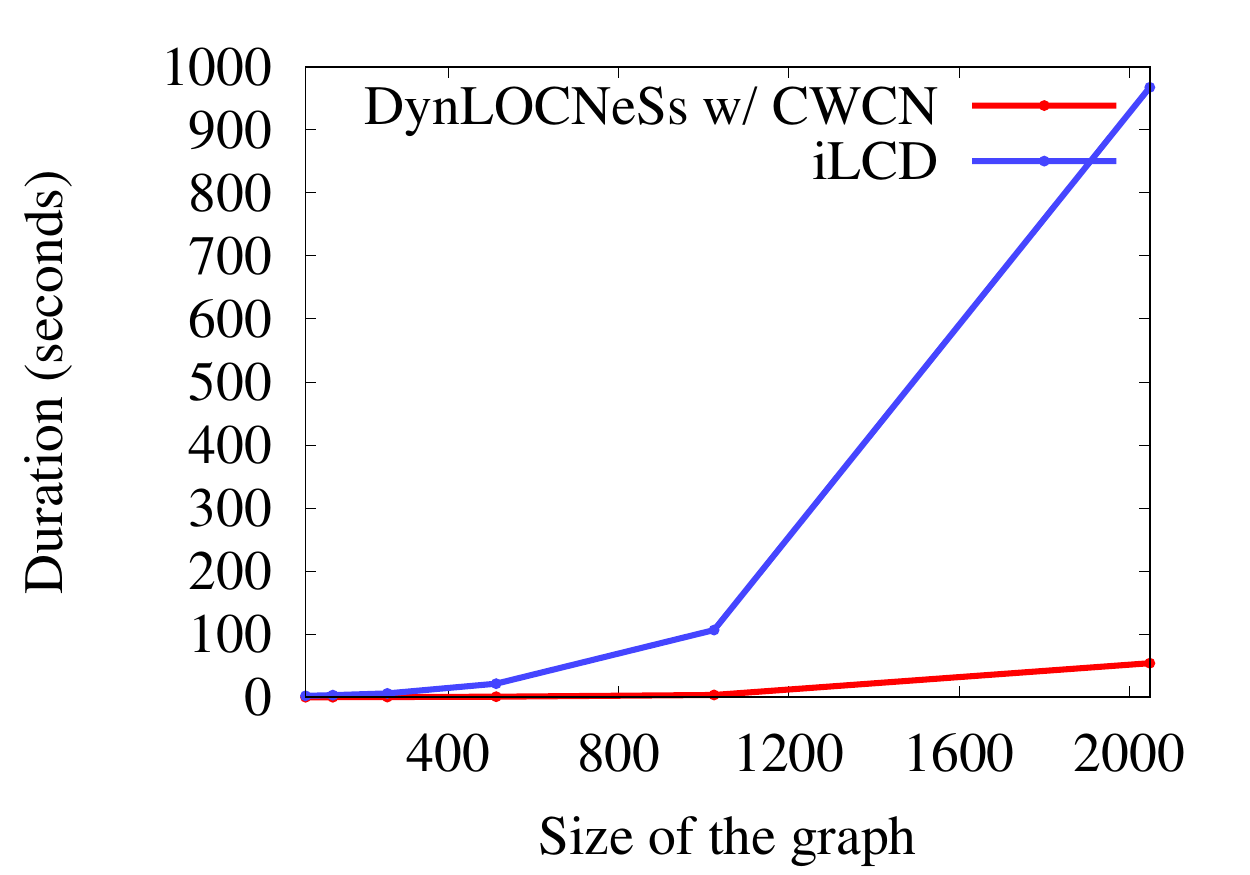}
		\caption{\small Speed of execution as a function of the graph size}
		\label{fig:exec}
\end{figure*}
Because the input and output of dynamic community detection algorithms depend on the dynamicity model used, it is difficult to compare to them. For example, \textit{iLCD} input is event-based (edge addition or deletion) and its output is a chronological sequence of community states. A state change can happen any time an edge is removed. The consequence is that, if launched on a time step sequence similar to those used to test the proposed algorithm, the community structure can vary several times during a same time step. Any heuristic to gather all the changes made during a time step would inevitably erase information and introduce a bias.

Another example, the multi-step adaptation of Louvain algorithm \cite{aynaud_static_2010} takes a sequence of time steps into account, but outputs a unique community structure at the end of the process and it is not possible to track the evolution of this structure during the detection process.

A more neutral comparison axis is the execution time, presented below, chosen to illustrate the performance of the proposed method : six graphs, of 64, 128, 256, 512, 1024 and 2048 vertices respectively, were generated with the same density as in the previous experiments: 0.05 intra-community and 0.5 inter-community, over 10 time steps. For example, the 1024 had roughly 375,000 edges to process over the 10 time steps.

We measure the mean time, over 5 runs, taken by \nomalgo{} and by iLCD to process each graph. The platform used is a Intel Core i7-2600K CPU @ 3.40GHz Workstation with 16GB RAM.

Results are presented on Figure \ref{fig:exec}. We can see that iLCD processing time is skyrocketing before the method we propose, which is a significant advantage to process either large graphs or large number of time steps.

\section{Conclusion and Future Works}
\label{conclusion}
We propose a new dynamic community detection method, named \nomalgo{} that consists in a vertex-centred approach to re-compute only a small local fraction of vertex neighbourhood. The algorithm relies on a vertex neighbourhood preference measure. We introduced a novel one, CWCN. Experiments on benchmark graphs show that CWCN yields better results than the other measures and that the overall method is well able to detect common patterns in community evolution such as grow-shrink and merge-split.

We are considering additionnal work on the community evolution patterns to better capture the dynamics and improve the quality of \nomalgo{} pattern identification. We are also working on experiments to assess the performance of the method on large graphs (up to millions of vertices).




\end{document}